\newcommand{\bfk}{\mbox{\boldmath {$k$}}}
\newcommand{\bfq}{\mbox{\boldmath {$q$}}}
\newcommand{\bfp}{\mbox{\boldmath {$p$}}}

\newcommand{\fslash}[1]{{\mbox{$\!\not\!#1$}}}

\newcommand{\mPi}{{\mit \Pi}}

\newcommand{\mSigma}{{\mit \Sigma}}

\newcommand{\refer}[1]{(\ref{#1})}

\documentstyle[12pt]{article}

\begin{document}

\setlength{\baselineskip}{0.2in}

\begin{titlepage}

\hspace*{12.7cm} HUPD-9511

\hspace*{12.7cm} March, 1995

\vskip 1.0cm
\Large
\centerline{\bf Fermionic excitation in quark
matter}

\vskip1.0cm

\normalsize
\centerline{Kazuhiro Tanaka}
\vskip0.2cm
\centerline{\it Department of Physics, Hiroshima University,
Higashihiroshima-shi, Hiroshima 724,
Japan\footnote{E-mail: tanakak@theo.phys.sci.hiroshima-u.ac.jp}}

\vskip2.0cm
{\bf Abstract:}
We discuss the fermionic excitation spectra of the
dense QCD and QED plasma at zero temperature,
and analyze the stability of the normal (perturbative)
ground state.
The standard re-summed
perturbation theory breaks down because the
infrared
(IR) divergences show up in the spectra
near the Fermi surface. We employ the
effective field theory approach and the renormalization
group to investigate the dynamics near the Fermi surface.
Our results indicate that the dense QCD and QED plasma
are non-Fermi-liquids.

\vskip6.0cm

Talk presented at YITP Workshop on

 ``{\it From
Hadronic Matter to Quark Matter:
Evolving View of Hadronic Matter}'',

YITP, Kyoto Japan, October (1994).

\vskip0.3cm
(to be published in {\it Supplement of Progress of Theoretical Physics})

\end{titlepage}

\newpage
\noindent
{\bf 1. Introduction}

There are several situations where hot and/or dense
deconfined quark-gluon plasma (QCD plasma) might be formed;
e.g., the interior of the neutron stars, the early universe,
and the relativistic heavy-ion collisions. It is commonly
accepted that for the QCD plasma the free gas
description gives a useful starting approximation.
The another example for which such treatments are frequently
used is the QED plasma (electron gas).
This corresponds to the following assumption: for
sufficiently hot and/or dense case, perturbation theory
(PT) along with some appropriate resummation procedure
(re-summed perturbation theory, rsPT) gives a successful
scheme to describe the dynamics.

In this talk we concentrate on the dense case, by which we
mean the high-density but zero-temperature case.
For this case, the above assumption corresponds to that the
ground state is the normal Fermi-liquid state.
We discuss
the fermionic excitation spectra in the dense plasma,
and examine the validity of the
assumption.
We emphasize that the physics of
the dense QCD and QED plasma is profoundly connected with the
IR behavior of the theory.

\vskip0.5cm
\noindent
{\bf 2. Perturbation theory calculation of the ground-state energy}

A peculiar property of QCD
is the asymptotic freedom, which is a basis of
successful applications of perturbative methods to
high-energy processes\cite{mu}.
Naively, one would expect that the asymptotic freedom would
support PT starting from the free quark gas.
Such a picture was pushed forward by Collins and
Perry\cite{co}. They employed a renormalization group (RG)
argument similar to those in the high-energy process. The
following behavior of the thermodynamic quantities,
when the Fermi momentum $p_{F}$ is scaled up by $\kappa$,
was obtained:
\begin{equation}
X(\kappa p_{F}, g, m, \mu) = \kappa^{D_{X}}X(p_{F},
\overline{g}(\kappa), \overline{m}(\kappa), \mu),
\label{eq:x}
\end{equation}
where $X$ stands for some thermodynamic quantity like the
energy density, and $D_{X}$ is its mass
dimension. $g$ and $m$ are the gauge coupling constant and the
quark mass renormalized at the scale $\mu$, while
$\overline{g}(\kappa)$ and $\overline{m}(\kappa)$ are the
corresponding running quantities. Because
$\overline{g}(\kappa) \rightarrow 0$ as $\kappa \rightarrow
\infty$, \refer{eq:x} seems to support the naive expectation
discussed above.

There is a subtle point in this sort of
RG arguments: the difficulties might
arise from the IR divergences, because the gluons are
massless and also $\overline{m}(\kappa) \rightarrow 0$ as
$\kappa \rightarrow \infty$.
In the medium, however, the gluons acquire the Debye screening mass
and the plasma mass due to the interaction with the quark
gas\cite{be}. Collins and Perry argued that these ``effective
masses'' would act as an IR cut-off.

Following this line, and in order to demonstrate explicitly IR
safety of PT at high density, the calculation of the
ground-state energy, up to and including the effects of
$O(\alpha^{2})$ ($\alpha = g^{2}/4 \pi$), was carried out
in \cite{ma,bal}.
At $O(\alpha^{2})$, the diagrams, which involve the 3-, 4-point
gluon couplings or the ghost-gluon coupling, contribute.
As for the IR divergences, their results
can be summarized as follows:\footnote{The ultraviolet (UV)
divergences can be completely removed by the counter terms
determined at $p_{F}=0$.
}
the second order ($O(\alpha)$)
graph is free of the IR divergences. At the fourth order
$O(\alpha^{2})$, the individual diagrams contain the IR
divergences, but most of them cancel out between the
different diagrams. The only one exception is the diagram
involving the two fermion bubbles; thus the order
by order PT is suffered from the IR divergences. Fortunately,
however, the resummation of the higher
order graphs of the RPA-type (ring diagrams)
renders the result finite. This is similar
to the well-known nonrelativistic (NR) electron gas
calculation\cite{ge}. The resummation of the RPA produces a
finite static screening mass to the ``dangerous modes'' of
the gluons, and the difficulty can be avoided.
The rsPT for the dense QCD plasma is IR safe in accord with
the conjecture of ref.\cite{co}, at least for the computation
of the ground-state energy.

There is one price due to the resummation: The re-summed RPA
series (the sum of the ring diagrams
involving two or more fermion bubbles)
give the contribution of
$O(\alpha^{2}\log\alpha)$, which are dominant compared with
the other $O(\alpha^{2})$ graphs.

\vskip0.5cm
\noindent
{\bf 3. Fermionic excitation spectra}

In this section we compute the quark excitation spectra in
the dense QCD plasma in the framework of rsPT discussed in sect.2.

The excitation energy of the quarks in the medium can be
defined as the change in the total energy when a quark is
added (removed) into (from) the system. This can be
expressed as a functional derivative of the energy density
${\cal E}$ with respect to the Fermi distribution function
$n(p)$:
\begin{equation}
\epsilon(p) = \frac{\delta {\cal E}}{\delta n(p)},
\label{eq:e}
\end{equation}
$\epsilon(p)$ is nothing but the quasiparticle energy in the
Landau Fermi-liquid theory\cite{lan}. In PT,
the (free) quark propagator in the rest frame
of the $T=0$ plasma is\cite{be}
\begin{equation}
S(p) = \frac{1}{\fslash{p} -m + i\varepsilon} + i \pi
\frac{\fslash{p} + m}{E(p)}\delta(p^{0} - E(p))n(p)
\equiv S_{F}(p) + S_{D}(p),
\label{eq:s}
\end{equation}
with $E(p) = \sqrt{\bfp^{2} + m^{2}}$ and $n(p) =
\theta(p_{F} - |\bfp|)$.
Because $S_{D}$ involves $n(p)$,
the functional derivative \refer{eq:e}
cuts the one quark-line in the diagrams
for the energy density. Including the contributions up to
$O(\alpha^{2}\log \alpha)$ for ${\cal E}$, we obtain
as the corrections to the free quark spectrum $E(p)$:
\begin{equation}
\epsilon_{1}(p) = \frac{m}{E(p)}\overline{u}(p) \mSigma(p)
u(p).
\label{eq:e1}
\end{equation}
$u(p)$ is the free quark spinor normalized as $\overline{u}u
= 1$. $\mSigma(p)$ is the RPA-type
self-energy given by the diagrams
of fig.1.
For the computation of $\epsilon_{1}(p)$, we
perform the Wick rotation for the gluon loop momentum
integration. The procedure is similar to the one discussed in
\cite{ta3,ta}. One obtains the two types of contributions,
depending on whether the
intermediate quark line connecting the external legs in the
diagrams of fig.1 is off-shell or
on-shell. We denote the off-shell part by
$\epsilon^{W}_{1}(p)$ while the on-shell part by
$\epsilon_{1}^{C}(p)$, as $\epsilon_{1}(p) =
\epsilon^{W}_{1}(p) + \epsilon_{1}^{C}(p)$.
$\epsilon^{C}_{1}$ can be compactly expressed as
\begin{equation}
\epsilon_{1}^{C}(p) = 2 N_{f}N_{c} \int^{p_{F}}_{p} \frac{dq \:
q^{2}}{(2 \pi)^{2}} \int^{1}_{-1} d\cos \theta f(\bfp,
\bfq).
\label{eq:e1c}
\end{equation}
$\cos \theta = \bfp \cdot \bfq/pq$, and
$f(\bfp, \bfq)$ is the exchange-type interaction (averaged
over spin and color) between
the two quarks with the momenta $\bfp$, $\bfq$.
Here and in the following, we consider the case of
SU($N_{c}$) color and the $N_{f}$-light quarks.

$\epsilon_{1}^{W}$ is given as a (Wick-rotated)
4-dimensional euclidean integral. It can be shown, by
explicit computation, that $\epsilon_{1}^{W}$ is free of the
IR divergences\cite{ta2}. On the other hand,
$\epsilon_{1}^{C}$ is IR divergent!
$f$ assumes the following form\cite{ta3,ta}:
\begin{equation}
f(\bfp, \bfq) = \sum_{X = L, T}\frac{A_{X}}{(p-q)^{2} -
\mPi_{X}(p-q)},
\label{eq:f}
\end{equation}
where ``L'' and ``T'' denote the longitudinal and the
transverse modes of the gluons. They are the two independent
gluonic modes in the medium. $\mPi_{X}$ is the self-energy
for these modes due to the RPA resummation. It is given by
the gluon self-energy $\mPi_{\mu \nu}$ as
$\mPi_{L}(k) = \mPi_{11}(k)-\mPi_{00}(k)$, $\mPi_{T}(k) =
\mPi_{33}(k)$ if we choose $\bfk$ to lie on the $x$-axis.
$A_{X}$ is the dimensionless function accounting for the
vertex structure of the quark-gluon coupling and is
proportional to $\alpha$.
The IR divergence shows up near the Fermi surface. For $p
\cong p_{F}$, we obtain
\begin{equation}
\epsilon_{1}^{C}(p) \cong (p - p_{F})\frac{N_{f}N_{c}}{4 \pi^{2}}
\sum_{X = L, T} \int^{1}_{-1} d\cos\theta \;\frac{A_{X}}{1 -
\cos\theta + \hat{\mPi}_{X}(p-q)},
\label{eq:e1c2}
\end{equation}
where the integrand is evaluated at the Fermi surface
($|\bfp|=|\bfq|=p_{F}$), and $\hat{\mPi}_{X}(k) =
\left(\mPi_{X}(k)/2
p_{F}^{2}\right)_{k^{0}=0}$.
The integral would be logarithmically
divergent if $\hat{\mPi}_{X}= O(\theta^{2})$ as $\theta
\rightarrow 0$. This is the IR divergence due to
the soft gluons with the momentum $p_{\mu}-q_{\mu}
\rightarrow 0$.

Now the key quantity which determine the IR behavior of the
fermionic excitation spectra is the so-called ``screening
mass'' of the gluons defined by
\begin{equation}
M^{2}_{X} = \lim_{k \rightarrow 0} \lim_{k_{0}\rightarrow
0}\mPi_{X}(\bfk, k_{0}).
\label{eq:m}
\end{equation}
In the RPA resummation scheme, $\mPi_{X}$ contains the quark
one-loop diagrams. One finds:
\begin{equation}
M_{L}^{2} = 2 N_{f}\frac{\alpha}{\pi}p_{F}^{2}; \;\;\;\;
M_{T}^{2} = 0.
\label{eq:ml}
\end{equation}
These are the gauge invariant results.
Therefore, the L-modes give the IR safe result, while the
T-modes cause the IR divergences. The screening of the L-modes
corresponds to the Coulomb
screening in the NR electron gas\cite{ge}. The
non-screening of the T-modes (magnetic modes)
is a result of gauge invariance $k_{\mu}\mPi^{\mu \nu}= 0$
in the framework of PT.

We now find for the quark excitation spectra:
\begin{equation}
\epsilon(p) = \epsilon_{F} + (p-p_{F})v + O\left( (p -
p_{F})^{2}\right),
\label{eq:ep}
\end{equation}
where $\epsilon_{F}$ is the Fermi energy, and
\begin{equation}
v = v_{0} - C_{F} \frac{\alpha(p_{F})}{\pi}v_{0}^{2}\left
[ \log\left| \frac{p_{F}}{p-p_{F}}\right| + O(1) \right],
\label{eq:v}
\end{equation}
with $v_{0}= p_{F}/E(p_{F}), C_{F} = (N_{c}^{2}-1)/2 N_{c}$.
$\alpha(p_{F})$ is the running
coupling constant at the scale $p_{F}$.
The result indicates the logarithmic singularity at $p =
p_{F}$, and thus the negatively infinite slope of the
spectra at the Fermi surface
($v_{F}=\partial \epsilon (p)/\partial p |_{p = p_{F}} \rightarrow
- \infty$).
The behavior of the spectra is shown in fig.2.
Several comments are in order here:

(a)
Also for the
case of the hot QCD plasma (at $p_{F} = 0$), the T-modes
of the gluons do not get screened,
at least up to $O(\alpha)$\cite{be}:
it was pointed out\cite{li}
that this causes the breakdown of PT for
the thermodynamic potential at the sixth
order. On the other hand, for the case of the dense
plasma, there has been no clear indication of such
difficulties: all energy density computations by rsPT
have been worked out successfully as discussed
in sect.2.
Our results reveal that
the RPA
breaks down if one computes the excitation spectra.

(b) In the present treatment,
the non-abelian character of the theory
does not enter explicitly, except for the running coupling
constant $\alpha(p_{F})$ and the color factors of
\refer{eq:v}. Thus the result can be
easily translated into the case of an abelian gauge theory,
i.e., into the QED plasma.
This indicates that
a naive extension of the famous work on the
NR electron gas by Gell-Mann and
Brueckner\cite{ge}
to the relativistic case breaks down:\footnote{The
ground-state energy of the relativistic electron gas was
obtained successfully in\cite{ak,ma,bal}} in the
NR case, the quasiparticle energy in the RPA
scheme was obtained successfully by Gell-Mann\cite{ge2}. Our
results \refer{eq:ep},\refer{eq:v}
do not reduce to Gell-Mann's result even if we go to the
NR limit ($p_{F}/m \rightarrow 0$). The
discrepancy comes from the IR divergent contributions
discussed above.
Our results confirm the argument of \cite{ba}
by employing the full relativistic framework.

(c) The breakdown of rsPT poses a question on
the usefulness of the free gas description of the dense QCD
and QED plasma. The spectra with the inversion may indicate the
instability of the normal (Fermi-liquid) state
(see also \cite{ba,se,ko,ta}).

The breakdown of the RPA as rsPT for high
density has a great impact.
Before drawing physical
implications, however,
we have to investigate the effects beyond the
RPA, and examine whether those effects cancel the above IR
divergences or not.
In the next section we employ the RG approach and try to go
beyond the RPA.
%

\vskip0.5cm
\noindent
{\bf 4. Renormalization group approach}

In this section we investigate the ground state of the
dense QCD plasma in the framework of the renormalization group
(RG) \`{a} la Wilson. The Wilson RG is a useful tool to analyze
the low energy dynamics of the system and to construct an
effective field theory.
It is especially powerful when the low energy
dynamics is dominated by gapless (massless) excitations. In
this case, the effective field theory
corresponds to an IR fixed point of
the RG transformation. A well-known example is the
application to critical phenomena\cite{wil}.

Recently, there has been progress in the
applications of the RG to (NR) many-fermion
systems\cite{sh,pol,ben,rho}.
The key observation is that the system of many-fermions in the
normal state
is dominated by gapless excitations at low energy, due to
particles and holes near the Fermi surface.
The Landau Fermi-liquid theory corresponds to the Gaussian
(mean-field) fixed point.
This view point also opens a possibility to classify the
possible phases of the system by a few number of ``relevant
perturbations'' which would lead to a RG flow toward another
IR stable fixed point rather than the Fermi-liquid fixed
point. For example, it has been shown\cite{sh,pol,ben}
that, for the case of
3-dimensional NR Fermi systems with
rotational invariance, only the pairing interactions
leading to the Cooper pairs could be (marginally) relevant;
otherwise,
the system is described as a normal Fermi-liquid.

Now we apply the RG approach to the dense QCD plasma, and
investigate whether the Fermi-liquid fixed point is a IR
stable one or not. If it were not stable, the assumption
discussed in sect.1 would be wrong.
As the first step,
we have to write down the low
energy effective action $S_{\Lambda}$. It is obtained by integrating out
the ``fast modes'', which have the momenta greater than the
UV cut-off $\Lambda$, in the original action. The modes
within $\Lambda$, which we call the ``slow modes'', constitute
our ``physical'' degrees of freedom dominating the low energy
dynamics, and $S_{\Lambda}$ should be expressed by
them. One important point is that all possible relevant or marginal
interactions (perturbations) between the slow modes, except those forbidden
by symmetry, should be included in $S_{\Lambda}$.
$S_{\Lambda}$ for our RG analysis is:
\begin{eqnarray}
S_{\Lambda} &=& \int_{|p-p_{F}| < \Lambda}d^{4}p\;
\overline{\psi}(p)\gamma^{0} \left[p^{0} - \epsilon_{F} - v_{F} (p -
p_{F}) \right]
\psi(p) + \int_{|q| < \Lambda}d^{4}q \; q^{2}
\left(\widetilde{A}^{a}_{i}(q)\right)^{2}
\nonumber \\
&+& \int_{|p'-p_{F}| < \Lambda}d^{4}p'
\int_{|p-p_{F}|<\Lambda}d^{4}p \int_{|q|<\Lambda}d^{4}q \; (2
\pi)^{4} \delta^{(4)}\left(p'-p-q\right) g
\overline{\psi}(p')\gamma^{i}t^{a} \psi(p) \widetilde{A}^{a}_{i}(q)
\nonumber \\
&+&({\rm \mbox{4-fermion interaction terms}}) + ({\rm \mbox{gluon
interaction terms}}).
\label{eq:sl}
\end{eqnarray}
Here the quark ($\psi$) and the gluon ($\widetilde{A}^{a}_{i}$)
fields are in the momentum
space.\footnote{These fields stand for the fermionic and the
bosonic elementary excitations with the same quantum numbers
as the quark and the gluon; we call them simply by the quark
and the gluon in this section.}
The momentum integrations are restricted within
$\Lambda$ and involve the
slow modes only. Note that, for the fermions,
the slow modes are those near the Fermi surface. Their
excitation spectra are linearized as $\epsilon(p) \cong
\epsilon_{F} + v_{F}(p - p_{F})$.
$\widetilde{A}^{a}_{i}$ stand for the T-modes of the gluons.
The L-modes of the gluons
do not appear in \refer{eq:sl}: as we saw in sect.3,
the ``mass'' of the gluons relevant for the fermionic
excitation spectra is the screening mass.
The L-modes become massive due to the Debye screening, and thus
decouple from the low energy dynamics. On the other hand,
the screening mass for the T-modes is forbidden by gauge
invariance. The integration of the L-modes and
other fast modes generate the 4-fermion interactions, but only a
few types of them, which satisfies special kinematical conditions,
could be marginal (at the tree level).
The ``gluon interaction terms'' may involve the 3- and
4-point couplings between the gluons, and also the gauge
terms (the gauge fixing terms and the Faddeev-Popov ghost
terms), but the explicit forms of them are irrelevant for the
following discussion.

As a result of
mode elimination, the coupling parameters ($v_{F}$, $g$,
the 4-fermion couplings, etc.) depend on $\Lambda$, and thus are the
running couplings in the RG. The partition function is
given as the functional integral by the slow modes
\begin{equation}
Z = \int \left[ d\psi d\overline{\psi}d \widetilde{A} \right]
e^{iS_{\Lambda}}.
\label{eq:z}
\end{equation}
We perform the RG transformation and derive the RG
equations for the running couplings. We divide our physical
space further into the slow modes ($|p-p_{F}|, q <
\Lambda/s$, $s>1$) and the fast modes ($\Lambda/s<
|p-p_{F}|,  q <\Lambda$), and integrate the fast modes out
in \refer{eq:z}. The scale
transformation $|p-p_{F}|, q \rightarrow |p-p_{F}|/s, q/s$
recovers the phase space reduced by mode elimination.
The comparison of the new effective action with
\refer{eq:sl} gives a flow of the coupling parameters
when $\Lambda$ is scaled
down. In order to trace the RG flow toward the Fermi-liquid
fixed point, we perform the scale transformation to the
quark and the gluon fields as
$\psi \rightarrow s^{3/2}\sqrt{Z_{2}}\psi; \;
\widetilde{A}^{a}_{i}
\rightarrow s^{3}\sqrt{Z_{3}}\widetilde{A}^{a}_{i},$
where $Z_{2}$ and $Z_{3}$ are the renormalization constants
for these fields.

The Fermi velocity $v_{F}$ determines the behavior of the
fermionic excitation spectra near the Fermi surface. This is
a marginal coupling at the tree level. By performing mode elimination
at the 1-loop level (see fig.3),
we obtain the RG equation:
\begin{equation}
\Lambda \frac{d v_{F}}{d \Lambda} = C_{F} \frac{\alpha}{\pi}
v_{F}^{2},
\label{eq:rge}
\end{equation}
with $\alpha = g^{2}/4\pi$.
This should be combined with the $\beta$-function, $\beta =
\Lambda \left( d\alpha/d\Lambda \right)$.
As a boundary condition, we match our effective field theory
with the QCD for the zero density, at the initial
(high-energy) scale $\Lambda_{0} (\sim p_{F})$ of
the successive RG transformation toward the Fermi surface:
$\alpha(\Lambda) = 1/\left[\beta_{0}\log(\Lambda/\Lambda_{{\rm
QCD}})\right]$, or $\beta= -
\beta_{0} \alpha(\Lambda) + O(\alpha^{2})$,
for $\Lambda \ge \Lambda_{0}$. ($\beta_{0} = (11N_{c} -
2N_{f})/6\pi$ and $\Lambda_{{\rm QCD}}$ is the QCD scale
parameter.)
This implies, by the continuity,
$\alpha(\Lambda)$ grows at the beginning
if we scale down $\Lambda$
from $\Lambda_{0}$. When we go to the lower energy, we
have the two possibilities, depending on the behavior of
$\beta(\alpha)$ for $\alpha \geq \alpha(\Lambda_{0})$:

(i) $\beta(\alpha) < 0$ (i.e., no zero of $\beta(\alpha)$) for
$\alpha \geq \alpha(\Lambda_{0})$.

(ii)
$\beta(\alpha^{\ast}) = 0$ for some $\alpha^{\ast} \geq
\alpha(\Lambda_{0})$.

The case (i) corresponds to the case where $\alpha$ is a
marginally
relevant coupling and continues to grow at low energy.
In this case, the low energy effective theory is a strong
coupling theory and thus does not correspond to the
Fermi-liquid fixed point.

The case (ii)
corresponds to the case where $\alpha^{\ast}$ gives an IR
fixed point. If $\alpha^{\ast}$ is small,
this may correspond to the Fermi-liquid
fixed point, and thus needs more detailed analysis.
In this case, \refer{eq:rge}
can be easily integrated
to give
\begin{equation}
v_{F}(\Lambda) = \frac{v_{F}(\Lambda_{0})}{1 +
C_{F}\left(\alpha^{\ast}/\pi\right)
v_{F}(\Lambda_{0})\log\left(\Lambda_{0}/\Lambda\right)}.
\label{eq:sol}
\end{equation}
Here $v_{F}(\Lambda_{0}) >0$.
The result implies $v_{F} \rightarrow 0$
when one goes to the low energy
limit $\Lambda \rightarrow 0$, i.e., near the Fermi surface.
Note that if we expand \refer{eq:sol} perturbatively and
retain the leading term, we recover the previous result
\refer{eq:v}. This indicates that the discussion of sect.3
corresponds to assuming the case (ii);
the RG makes it possible to sum up the leading
logarithms in all orders of PT.

One might say that the result is ``safe'' if the case (ii)
is realized. We
have no IR difficulty. \refer{eq:sol} indicates that $v_{F}$ is
a marginally irrelevant coupling at low energy. So,
the low energy effective theory would be given by
the Fermi-liquid fixed point with $\alpha = \alpha^{\ast}$ and
$v_{F} = 0$. However,
we argue that this fixed point
will never be stable:
first, because $\alpha^{\ast} \neq 0$, there exist the
long-range attractive
interaction between the quarks due to the
exchange of the T-modes of the gluons. Second,
the density of states at the
Fermi surface $N(0) \propto 1/v_{F}$ is infinitely large
(see fig.2).\footnote{In the NR case
where the T-modes of the gauge fields are not included, the
running of $v_{F}$ is not so significant, and $v_{F}$ is
considered to go to a nonzero value for $\Lambda \rightarrow
0$\cite{sh}.}
The infinite
density of states will enhance the attractive channels of
the 4-fermion interactions, and will drive the instability
of the Fermi-liquid fixed point.
For example, the BCS-type paring interaction $V$
gets the 1-loop correction $\sim - N(0)V^{2}$, and is
strongly enhanced at low energy for $V<0$.
As an origin of the attractive $V$, we need not other
degrees of freedom like the phonon in the usual
superconductors; it could be
generated solely through mode elimination in gauge theory, e.g.,
by the Kohn-Luttinger effect\cite{kl}.

Our RG analysis indicates the two possibilities for the
effective field theory for the dense QCD plasma: (i) a
strong coupling theory with the large $\alpha$; (ii) a
``would-be fixed point'' with $\alpha = \alpha^{\ast}$, $v_{F}=0$.
We conclude that the ground state of dense QCD plasma is not
the normal Fermi-liquid state for both cases.

In order to determine which case is realized, we have to
compute the $\beta$-function from the relevant 1-loop
diagrams.
For the simpler case of the dense QED plasma,
we can show it corresponds to (ii) at
1-loop\cite{ta2}. (Note, in this case, $\beta_{0} <0$.)
Thus, the results similar to \refer{eq:sol} can be obtained:
the ground state of the
electron gas may be the superconducting state, even
without the phonon degrees of freedom.\footnote{The
possibility that the ground state of the relativistic
electron gas would be the superconducting state was
discussed by Chu, Huang and Polonyi by a different
argument\cite{chu}.}

The behavior $v_{F}(\Lambda)\rightarrow 0$ ($\Lambda
\rightarrow 0$) is similar to that of a marginal
Fermi-liquid.\footnote{The $v_{F}\rightarrow 0$ behavior has
been also found in a 2-dimensional Fermi system with a Chern-Simons
gauge field by Nayak and Wilczek\cite{na}.} The marginal
Fermi-liquid is a successful phenomenological model proposed
to explain the puzzling normal state properties of Cu-O
high-temperature superconductors\cite{var}.
That a proper relativistic treatment of the
electron gas might lead to superconductivity of a marginal
Fermi-liquid would be an interesting possibility as a model
of high-temperature superconductivity.

%
\vskip0.5cm
\noindent
{\bf 5. Conclusion}

We discussed the fermionic excitation spectra of the
dense QCD (and QED) plasma. It is shown that the standard rsPT
breaks down because the IR
divergences show up in the spectra near the Fermi surface.
The divergence is due to
the non-screening of the transverse modes of the gauge
fields.
By employing the effective field theory approach and the RG,
we classified the two possibilities for the
ground state of the QCD plasma:
(i) a strong coupling theory where the nonperturbative effects
will be important; (ii) a would-be IR fixed point with
$\alpha = \alpha^{\ast}$ and $v_{F}=0$, which would be unstable
against the transition to the superconducting state.
We concluded that the $T=0$ QCD (and QED)
plasma is a non-Fermi-liquid.

We stress that the physics of the dense QCD and QED
plasma is not so simple on the contrary to the usual picture.
Our results indicate the importance of the ``soft degrees of
freedom''; they could appear in the higher order terms in PT
as well as in the nonperturbative effects. More efforts are
needed to reveal roles of the soft degrees of freedom and
to understand the dynamics of the QCD and QED plasma.


\newpage
\centerline{\bf Figure Captions}

\vskip 15pt
\begin{description}

\item[Fig. 1] The self-energy diagrams for the quarks in the
RPA (full lines: quarks; wavy lines: gluons).

\item[Fig. 2]
The fermionic excitation spectra (dotted line: $v_{F} >0$;
dashed line: $v_{F} = -\infty$; full line: $v_{F} = 0$).

\item[Fig. 3]
The 1-loop diagrams which give the RG equation \refer{eq:rge}.
The first diagram is due to the transverse gluon exchange,
while the second one is due to the 4-fermion interactions.

\end{description}

\end{document}